# Charge Transport in Semiconductors Assembled from Nanocrystals

Nuri Yazdani[1], Samuel Andermatt[2], Maksym Yarema[1], Vasco Farto[1], Mohammad Hossein Bani-Hashemian[2], Sebastian Volk[1], Weyde Lin[1], Olesya Yarema[1], Mathieu Luisier[2], Vanessa Wood[1]*

**Affiliations:**

[1]Materials and Device Engineering Group, Department of Information Technology and Electrical Engineering, ETH Zurich, Zurich CH-8092 Switzerland.

[2]Nano TCAD Group, Department of Information Technology and Electrical Engineering, ETH Zurich, Zurich CH-8092 Switzerland.

*Correspondence to: vwood@ethz.ch

The potential of semiconductors assembled from nanocrystals (NC semiconductors) has been demonstrated for a broad array of electronic and optoelectronic devices, including transistors, light emitting diodes, solar cells, photodetectors, thermoelectrics, and phase charge memory cells. Despite the commercial success of nanocrystals as optical absorbers and emitters, applications involving charge transport through NC semiconductors have eluded exploitation due to the inability to predictively control their electronic properties. Here, we perform large-scale, ab-initio simulations to understand carrier transport, generation, and trapping in NC-based semiconductors from first principles. We use these findings to build a predictive model for charge transport in NC semiconductors, which we validate experimentally. Our new insights provide a path for systematic engineering of NC semiconductors, which in fact offer previously unexplored opportunities for tunability not achievable in other semiconductor systems.

**Main Text**

Assembly of colloidal nanocrystals (NCs) into thin films[1] is envisaged as a means to achieve next generation, solution-processed semiconductors with electronic properties (e.g., band-gaps,[2] band-edge positions,[3] mobilities,[4] and free carrier densities[5]) that can be defined to match specific application requirements.[6–10] This tunability is enabled by a multi-dimensional design space, where size, shape, composition, surface termination, and packing of the NCs can be systematically and independently controlled. While parametric studies have demonstrated some of the scaling relations in this design space,[11–15] the fundamental mechanism driving charge transport in NC-based semiconductors has remained unclear, making it difficult to build up predictive models for charge transport in NC semiconductors or tap the full potential of NCs as building blocks for electronic materials through theory-guided design.

Here, we perform large-scale, *ab-initio* simulations to understand carrier transport, generation, and trapping in NC-based semiconductors from first principles. We use these findings to build and experimentally validate a predictive model for charge transport in NC semiconductors. This predictive model allows us to design NC semiconductors with unique properties not achievable in the bulk, and the fundamental insights into charge carrier dynamics sets a clear agenda for the development of NC chemistry and self-assembly to realize novel semiconductors.

### *Ab initio* **Investigation of Charge Transport**

Only recently has it been computationally feasible to treat the full atomic complexity of a NC *ab initio*, and this has proven key to understanding the mechanisms driving charge carrier dynamics on individual NCs.[16] In order to elucidate the mechanisms for charge transport in NC-based semiconductors, we implement large-scale density functional theory (DFT) calculations on individual NCs, and on systems containing up to 125 NC (> 200000 atoms) (**Fig. 1a-b**).

As a model system, we use lead sulfide (PbS) NCs in the quantum confined regime (i.e., with radii $r < \sim 3$ nm) terminated with iodine ligands.[17] Since NC size is a parameter that is easy to systematically control in experiments, we perform calculations for different sized NCs in order to validate the resulting charge transport model with experiments. Details are provided in the **Methods**.

*Polaron Formation and Reorganization Energy*

Before we understand how charge moves across a NC-based semiconductor, we must first consider the impact of the presence of a charge carrier on an individual NC. To do so, we compute the ground state physical structure of the NCs in their neutral charge state and when charged. Upon charging with an electron (or hole), the Pb-iodine ligand bonds on the (111) surfaces of the NCs expand (or contract), while the Pb-S bond lengths remain unchanged (**Fig. 1c, Fig. S1,** and **Supplementary Information note 1**). Thus, the presence of a charge carrier on a NC can lead to the formation of a polaron. Because such polarons result from electrostatic interaction of the charge carrier with the negatively charged functional group of the ligands, polaron formation can be expected in any NC system with X-type ligands[18] (e.g. halides, thiols, carboxylates).

Charge transfer from one NC to another thus implies a rearrangement of atoms at the surface of the two NCs. Although the shifts in bond length are small (up to 0.5 pm or 0.02% of the 3.22Å nominal bond length), the associated reorganization energy for charge transfer between two NCs, $\lambda$, is large (10s to 100s of meV) (**Fig. 1d**). The reorganization energy decreases with increasing NC size due to a reduced carrier density across the NC and an increased number of ligands. In the **Supplementary Information note 2**, we explain why it is reasonable to ignore the contribution to $\lambda$ stemming from reorganization of the neighboring NCs (i.e., outer-shell reorganization).

*Electronic Coupling in NC Semiconductors*

Having now understood that the presence of a charge carrier on a NC can lead to the formation of a polaron, we can determine the type of transport (band-like or hopping) by calculating the electronic coupling between neighboring nanocrystals, $V_{ct}$.

Small angle x-ray scattering measurements on PbS NCs have demonstrated that they assemble into a body-centered-cubic (BCC), face-center-cubic, and related structures, along with alignment of the individual NCs with respect to the superlattice structure as depicted in **Fig. 1e** for a BCC structure.[19–21] We consider two relative orientations for neighboring NCs: [111]-neighbors and [100]-neighbors (**Fig. 1e**). We calculate electron and hole couplings for both orientations (i.e., $V_{111}$ and $V_{100}$) over a range of $r$ and inter-NC facet-to-facet distances, $\Delta_{ff}$ (**Fig. S2** and **Supplementary Information note 3**). The results for $\Delta_{ff} = 6$Å are shown in **Fig. 1d**. $V_{ct}$ increases strongly as the size of the NC decreases in agreement with analytical calculations modelling the NCs as spherical potential wells.[22] This trend is explained by an increased carrier density on the outer atoms of the NC with increasing confinement in smaller NCs. The coupling in the [100] direction is about an order of magnitude larger than in the [111] direction, due to strong confinement of the carriers away from the ligand-rich [111] facets.[16]

*Phonon-Assisted Charge Transfer*

The fact that $V_{ct}$ is more than an order of magnitude smaller than $\lambda$ over the range of $r$ and $\Delta_{ff}$ of typical PbS NC-semiconductor with X-type ligands informs us that, in these systems, the charge carriers are polarons localized to individual NCs, and that charge transport occurs through a phonon-assisted charge transfer (polaron hopping) between neighboring NCs.

Since the charge carrier deforms the Pb-ligand bonds upon polaron formation, charge transfer will be driven by the Pb-ligand vibrations. *Ab-initio* calculations of the phonon density-of-states of PbS NCs,[16] backed by inelastic neutron[23] and x-ray[24] scattering, indicate that Pb-ligand vibrations for common X-type ligands occur at energies $\hbar\omega < {\sim}15$ meV. Therefore, at temperatures above ~175K, charge transfer will occur at a rate:[25]

$$k_{ct} = N_P \frac{2\pi}{\hbar} V_{ct}^2 \sqrt{\frac{1}{4\pi\lambda k_B T}} e^{-(\Delta E + \lambda)^2/4\lambda k_B T}, (1)$$

where $\Delta E = E_P - E_R$ (i.e. the energy of the products minus the energy of the reactants) and $N_P$ is the number of degenerate product states (**Fig. 1f**). At temperatures below ~175K, transfer rates will saturate to their temperature-independent, low temperature limit (**Supplementary Information note 4**), or, in the presence of disorder, transport will transition to an Efros-Shlovskii variable range hopping regime.[26]

Assuming a NC-semiconductor of isoenergetic NCs and no applied field ($\Delta E = 0$), **Eq. 1** predicts charge transfer times on the order of 10-100s ps for PbS NC-semiconductors at room temperature, in agreement with recent measurements.[13] Since intra-band carrier cooling rates in PbS NCs proceed at 100s fs times scales,[27] the reactant and product states are thus the highest occupied electronic states (in the case of hole transport) or the lowest unoccupied electronic states (in the case of electron transport). This is in agreement with experimental measurement of the mobility band gap in a NC-semiconductor scaling linear with the band gap of the individual NCs.[28]

*Energetic Landscapes in NC Semiconductors*

In a realistic NC-semiconductor, $\Delta E \neq 0$, with differences in the alignment between the highest occupied (or lowest unoccupied) states of neighboring NCs contributing to $\Delta E$. Since a large $\Delta E$ will have significant impact on the time scales of transport, it is therefore critical to understand and control the energetic landscape within a NC solid. One contribution to $\Delta E$ is the distribution of the individual NC bandgaps, stemming from size and shape disorder of the constituent NCs. Additionally, deep, electronic trap states are known to exist in NC-semiconductors,[29] and possible explanations of their origin include mid-gap states on individual NCs[30] and fused NC dimers.[31] Here, we demonstrate oxidized or reduced doped NCs in the NC-semiconductor also form electronic traps in NC-semiconductors.

An individual NC is doped according to the oxidation-number sum rule:[32,33]

$$N_C V_C + N_A V_A + \sum_i V_i = \begin{cases} 0, & \text{intrinsic,} \\ < 0, & \text{p}-\text{doped,} \\ > 0, & \text{n}-\text{doped,} \end{cases} \quad (2)$$

where $N_x$ is the number of the cations (C) and anions (A), and $I$ are the impurities and ligands with valence $V_x$ comprising the NC. Doped NCs will in general be energetically unfavorable; however, small densities of doped-NCs are to be expected through reaction kinetics.[33] For PbS-NCs for example, an excess of Pb during synthesis can lead to a small fraction of *n*-doped PbS NCs, and exposure of PbS NC-semiconductors to oxygen leads to *p*-doping.

We compute the electronic structure of a NC-semiconductor containing a single *n*-doped NC surrounded by intrinsic NCs (**Fig. 2a**). For reference, the electronic structures of isolated intrinsic, *n*-doped, and oxidized *n*-doped NC (with net charge $e+$) are shown in **Fig. 2b**. In a NC semiconductor, if the *n*-doped NC is not oxidized, its energy levels remain aligned with those of the neighboring intrinsic NCs (**Fig. 2c**). However, oxidation causes a shift in the energy levels of the *n*-doped NC as well as its neighbors (**Fig. 2c**, **Fig. S3**).

Within a NC-semiconductor, oxidized *n*-doped NCs thus behave as electronic traps for electrons and as barriers for hole transport. Equivalently, reduced *p*-doped nanocrystals present traps for holes and barriers for electrons **(Fig. S4).** Defining the trap depth ($E_T(r, \Delta_{ff})$) as the extent of the shift of energy level in an oxidized or reduced doped NC relative to a NC infinitely far from the doped NC (**Fig. 2d, Fig. S5**), we find that our calculated trap depths agree with the experimentally measured trap depths[23,28] as well as the charging energy of the NCs computed using the measured size-dependent dielectric constant for PbS NC-semiconductors.[34] Thus, while trap states have been typically ascribed to mid-gap electronic states on individual NCs, traps will also be presented by the presence of charged, doped-NCs in a NC-semiconductor.

In this picture, trapping and release of charge carriers from traps is thus simply phonon-assisted charge transfer between the highest occupied or lowest unoccupied states of neighboring NCs with rates given by **Eq. 1**, where $E_T(r, \Delta_{ff})$ is included in $\Delta E$. Doing so results in release rates on the order of $10^2$ for $r = 1$ nm NCs and up to $10^8$ for $r = 3$ nm, in agreement with the rates characterized previously with thermal-admittance-spectroscopy[23] (see **Fig. S6**).

These results also indicate that the excess carrier on a doped NC must overcome a large energetic barrier (equal to $E_T$) to become a free carrier in the NC-semiconductor. Particularly in small NCs, where $E_T$ is large, free carrier densities in NC-based semiconductors will be negligible even when large densities of doped-NCs are present. For example, for a semiconductor made of $r = 1.6$ nm PbS NCs, assuming 1% of NCs are *n*-doped, the free electron density at room temperature will be $\sim10^{12}$ cm$^{-3}$, in stark contrast to the total density of *n*-doped NCs, $\sim10^{16}$ cm$^{-3}$. However, the formation of a space-charge region can lead to oxidation (or reduction) of the doped NCs, resulting in high trap densities.

To summarize, our calculations provide several key insights into charge transport in semiconductors assembled from NCs: 1) charge on individual NCs forms polarons, 2) charge transport occurs via phonon-mediated charge transfer, and 3) oxidized or reduced doped-NCs become electronic traps states within the NC-semiconductor.

**Experimental Validation of Charge Transport and Trapping Models**

We experimentally validate these insights into charge transport in NC-based semiconductors by performing time-of-flight (TOF) photocurrent transient measurements[4] (see **Methods**, **Fig. S7**). In a TOF measurement, a laser pulse generates a low density of charge carriers in the NC-semiconductor (**Fig. 3a**), and the displacement current generated by the electrons or holes traversing the film of thickness $d$ is measured for a range of biases across the device, $V_B$, at temperatures $T$ ~220 - 330K (**Fig. 3b**). The resulting transients can be fit with two distinct power laws at short and long times (**Fig. 3b**), with their intersection taken as an effective transit time, $t_{tr}(V_B,T)$. $t_{tr}(V_B,T)$ corresponds to the maximum of the statistical distribution of carrier transport times across the device, and, by fitting $t_{tr}(V_B,T)$ simultaneously for all temperatures $T$ and biases $V_B$ (**Fig. 3c**), it is possible to extract an effective mobility, $\mu_{eff}$:

$$\frac{d}{t_{tr}(V,T)} = \mu_{eff}(r, \Delta_{ff}, T)\frac{(V_B+V_{B0})}{d}, (3)$$

where $V_{B0}$ is the built-in field in the device. The long-time portion of the transient reflects the large dispersion in carrier transit times, and are discussed further in **Supplementary Information note 6**.

We first note that, in the limit that the potential drop across neighboring NCs is smaller than the reorganization energy (($V_B + V_{B0}$)($2r+ \Delta_{ff}$)/$d << \lambda$), we can use **Eq. 1** to write:

$$\mu_{eff}(r, \Delta_{ff}, T) = N_p \frac{2\pi}{\hbar} \frac{V_{ct}^2}{\lambda^{1/2}} \frac{(2r+\Delta_{ff})^2}{2k_BT} \sqrt{\frac{1}{4\pi k_BT}} e^{-E_A/k_BT}. (4)$$

and thereby extract $V_{ct}$ and $E_A$ from experiment (**Fig. S8**). We find good agreement between the experimentally extracted $V_{e*}$ and $V_{h*}$ and computed values for electron and hole coupling in the [100] direction (**Fig. 3d**). The extracted activation energies, which range from 70 meV to 150 meV (**Fig. 3e**), are larger than those expected for a NC-semiconductor with no energetic disorder ($E_A = \lambda/4$ ~ 10 meV to 40 meV) (**Fig. 1d**). Instead, the activation energies are consistent with the values expected when electronic traps dominate the timescales of carrier transport. The fact that the electronic coupling $V_{ct}$ measured in this trap-limited transport regime agrees with our calculations of $V_{ct}$ between neighboring NCs indicate that the trap states limiting the effective mobility in NC-based semiconductors are those stemming from oxidized (or reduced) doped-NCs.

**Predictive Model for Charge Transport**

Confident in our new understanding of charge-transport in NC-based semiconductor, we build a Kinetic Monte Carlo (KMC) simulation of polaron transport, which we parameterize with the DFT calculated values for electronic coupling $V_{ct}$, reorganization energy $\lambda$, and electronic trap depth $E_T$ (See **Methods**). In this multiscale model, charge transport across a NC-based semiconductor is simulated as sequential charge transfers between neighboring NCs $i$ and $j$. The rate of charge transfer is $k_{ij}$ (which is given by **Eq. 1**) with energy offset between neighboring NCs, $\Delta E_{ij}$, is taken as

$$\Delta E_{ij} = (E_{g,j} - E_{g,i})/2 - \vec{E}_z \cdot (\vec{r}_j - \vec{r}_i), (5)$$

where $E_{g,i}$ is the band-gap of NC $i$, $E_z$ is the electric field across the NC-semiconductor (assumed to be in the $z$-direction), and $r_i$ are the coordinates of NCs. For our simulation, we construct artificial NC semiconductors having the thicknesses and containing the different sized NCs that are investigated experimentally with TOF, and simulate current transients for different biases, carrier types (electrons and holes), and temperatures. Only the density of trap states as a function of NC size, $p_T(r)$, is left as a free parameter (**Supplementary Information note 7**). The examples shown in **Fig. 4a** highlight that all simulated transients (red lines) match the measured transients (blue lines), both the effective mobilities defined by $t_{tr}(V_B,T)$, as well as the long-time ($t > t_{tr}(V_B,T)$) dispersion of the transients. We find trap states densities selected to achieve agreement are within the expected range, and otherwise, no fitting is carried out to achieve the agreement between the simulated and experimental current transient measurements.

**Discussion**

This predictive, multiscale model can be used to systematically design next generation NC-based semiconductors. Here, we consider how to overcome one intrinsic limitation we identified, namely the one-to-one correspondence between the free carrier generation in a NC and the formation of deep traps.

In **Fig. 4b**, we plot the simulated relative mobility of a PbS NC-semiconductor, ($\mu_{eff}/\mu_0$), defined by the time required for ~63% of the carriers to traverse a 400 nm-thick film, as a function of bandgap disorder $\sigma_{Eg}$ and trap density $\rho_T$, relative to that for a trap- and disorder-free NC-semiconductor, $\mu_0$. While band-gap disorder has a similar impact on carrier mobility for both small and large NCs, the impact of deep traps can most easily be mitigated by using larger NCs, or by significantly decreasing $\Delta_{ff}$ (e.g. through epitaxially connected NC-semiconductors [11]) since $E_T(r, \Delta_{ff})$ decreases with increasing $r$ and decreasing $\Delta_{ff}$).

Our insights enable us to identify a more flexible approach: a NC semiconductor composed of intrinsic NCs can be doped with p- or n-doped NCs with bandgaps larger than that of the intrinsic NCs. With proper selection of bandgap, the shifted highest occupied state of oxidized *p*-doped NCs or the lowest unoccupied state of reduced *n*-doped NCs will align with the highest occupied states or lowest unoccupied states of the intrinsic NCs (**Fig. 4c**). This simultaneously eliminates deep traps and energetic barriers for thermal release of carriers, and leads to multiple orders of magnitude higher mobilities and free carrier densities. Such a strategy can be achieved with a bimodal size distribution of NCs or with equal-sized doped NCs of a different core material (and thus different bandgaps).

In summary, our insights highlight the need to reframe how we think about charge transport, trapping, and doping in NC semiconductors. As previously discussed, we should expect to find similarly large reorganization energies for any small NCs with X-type ligands, and the formation of trap states upon charging of doped NCs should similarly occur, and the approaches employed here can be readily adapted in order to parameterize other NC-semiconductors. Finally, we note that our modeling here assumes negligible exciton polarization across individual NCs. For large fields and/or weak confinement, both the reorganization energy and electronic coupling will become field dependent, effects will which will need to be accounted for in modelling of the transport.

Within the polaron hopping regime, NCs semiconductors present highly tunable systems that offer complete control of electronic coupling through tuning of the electronic confinement in the individual NC, the spacing, and the topology of the NC lattice, as well as the activation energies associated with transport through tuning of the NC dispersity, doping, and surfaces. By controlling the phonon densities-of-states and electron-phonon coupling through atomic engineering of the NCs and their surfaces, the rates and temperature-dependences of transport can also be systematically tuned. The example of PbS NC-based semiconductors illustrates how it is possible to engineer electronic anisotropies into semiconductors (i.e., transport in [100] will be faster than in [111]), without resorting to anisotropic crystal structures. This enables the creation of semiconductors with isotropic optical properties but with highly anisotropic electronic properties (as in the case of PbS), or with highly anisotropic optical properties and highly isotropic electronic properties. These findings position NC semiconductors not only as highly tunable, solution-processed semiconductors but also as model, tunable systems for studying the fundamental physics of charge transfer processes.

**Methods**

**Atomistic Model Construction for DFT Calculations**

Atomistic models for the NCs are constructed according to the atomistic model proposed by Zherebetskyy et al.[17] Bulk rocksalt PbS (with a Pb or S atom centered on the origin) is cut along the eight (111) planes and six (100) planes at plane to origin distances ($r$) defined by the Wulff ratio $R_W$

$$r_{(1,0,0)} = AR_W, \quad r_{(1,1,1)} = AR_W^1. \quad \text{(M1)}$$

A $R_W = 0.82$ is used.[17] The scalar $A$ is adjusted such that the resulting NC is S-terminated on the (111) facets. These (111)-surface terminating S atoms are then replaced with the desired ligand (for all calculations here, iodide anions). To obtain an intrinsic semiconductor NC, overall charge balance must be maintained according to eq. 4 in the main text. For all NC sizes investigated here, 1-2 ligands, or lead-ligand pairs are removed from the as cut atomistic model in order to satisfy charge balance. For the doped NCs, a single ligand or lead-ligand pair is additionally removed. All removed atoms are taken from the corners of the NCs corresponding to the intersection of the [111] and [100] facets.

**Electronic Structure and Electron Transfer Parameterization Calculations**

All electronic structure calculations are performed within the CP2K program suite utilizing the quickstep module.[35] Calculations are carried out using a dual basis of localized Gaussians and plane-waves,[36] with a 300Ry plane-wave cutoff. Double-Zeta-Valence-Polarization (DZVP),[37] Goedecker–Teter–Hutter pseudopotentials for core electrons, and the Perdew–Burke–Ernzerhof (PBE) exchange correlation functional are used for all calculations, as in previous calculations for PbS NCs.[16,23] Convergence to $10^{-8}$ in Self-Consistent Field calculations is enforced for all calculations unless otherwise specified.

Non-periodic boundary conditions in atomic coordinates and electric potential are used (with the exception of the superlattice calculations which uses periodic boundary conditions for both), through the use of a wavelet Poisson solver.[38] Geometry optimization is performed with the Quickstep module utilizing a Broyden–Fletcher–Goldfarb–Shannon (BFGS) optimizer. All atoms in all systems are relaxed using maximum force of 24 meVÅ-1 as convergence criteria.

**Reorganization Energy Calculations** Reorganization energies are calculated using a half- cell approach.[39] We first fully geometrically relax the atomic coordinates, $Q_x$, and compute the total energy of the neutral NC, $E_n(Q_n)$, the NC with an additional electron, $E_e(Q_e)$, and a NC with a hole, $E_h(Q_h)$. We then perform energy calculations, without any geometry optimization, for $E_n(Q_e)$, $E_n(Q_h)$, $E_h(Q_n)$, $E_e(Q_n)$. Then,

$$\lambda_{ih} = E_{Tot}(\Psi_P, Q_R) - E_{Tot}(\Psi_P, Q_P) = (E_n(Q_h) + E_h(Q_n)) - (E_n(Q_n) + E_h(Q_h)),$$
$$\lambda_{ie} = E_{Tot}(\Psi_P, Q_R) - E_{Tot}(\Psi_P, Q_P) = (E_n(Q_e) + E_e(Q_n)) - (E_n(Q_n) + E_e(Q_e)).$$
(M2)

*Electronic Coupling* These calculations are performed on a system of two NCs, oriented according to the two configurations in Fig. 2 of the main text, for all NC sizes, *r*, and a range of facet to facet separations, $\Delta_{ff}$. $V(r,\Delta_{ff})$ for the CBM and VBM is then taken as half the splitting of the resulting antisymmetric and symmetric states in the combined system.[40]

**Superlattice Electronic Structure Calculations** The electronic structure of the NC superlattices were calculated utilizing the Kim-Gordon method (KG), which partitions a weakly interacting system into subunits. Namely, it forces the overall Hamiltonian of the superlattice (the weakly interacting system) to be block diagonal, where each block corresponds to a strongly interacting subunit (each individual NC). The KG approximation should be very much suitable for the NC-superlattices, given our finding of weak electronic coupling between the NCs, and localization of charge carriers on individual NCs as a result of the large reorganization energies associated to polaron formation. A detailed description of the KG method can be found elsewhere.[41] For the calculations here, a linear-scaling approach to self-consistent field was employed, using a full embedding potential for the nonadditive kinetic energy correction to the PBE functional. Calculations are performed on 5x5x5 structures (corresponding to ~200 000 atoms for the largest NC), as calculations as a function of superlattice size indicate a convergence of the trap-depth with this size (see **Fig. S2**). In **Fig. S2** we additionally plot the trap depth for the 0.9 nm NC as a function of $\Delta_{ff}$, which, as expected, indicates an increase in trap depth with an increase in NC-NC separation resulting from weaker screening.

**Kinetic Monte Carlo Simulations**

*General* The Kinetic Monte Carlo simulations are performed within the limit of low charge carrier concentration, which assumes negligible interaction between the charge carriers. Charge transport is then simulated as sequential charge transfers (CT) between NCs. The CT rate between two NCs $i$ and $j$, $k_{ij}$, is given by Eq. 1 in the main text,

$$k_{ij} = N_P \frac{2\pi}{\hbar} V_{ij}^2 \sqrt{\frac{1}{4\pi\lambda k_B T}} \exp\left[-\left(\Delta E_{ij} + \lambda\right)^2 / 4\lambda k_B T\right], \text{(M3)}$$

where $\Delta E_{ij}$ is the energy of the products minus the energy of the reactants, and $N_P$ is the number of degenerate product states. We take $\Delta E_{ij}$ as

$$\Delta E_{ij} = \frac{E_{g,j} - E_{g,i}}{2} - E_z(\vec{r}_j - \vec{r}_i) \cdot \hat{z}, \text{(M4)}$$

where $E_{g,i}$ is the band-gap of NC $i$, $E_z$ is the electric field across the NC-solid (assumed to be in the $z$-direction), and $r_i$ is the coordinate of NC $i$.

For PbS NCs, intervalley coupling which stems from the [100] facets of the NC, break the 4-fold degeneracy of the valence band maximum (VBM) and conduction band minimum (CBM) of bulk-PbS in the NCs. This results in NC conduction band minima and valence band maxima of which are either singly or triply degenerate. This splitting is discussed in detail elsewhere [16], but we find the ordering to vary between differently sized NCs, i.e. for some sizes the VBM/CBM are singly/triply degenerate, whereas for other sizes, they are found to be triply/singly degenerate. We therefore use an average of $N_P = 2$ for both electron and hole transfer.

If a charge carrier is assumed to be on NC $i$, the NC to which it hops, and the time required for that hop to occur are determined in the following way [42]: First, the CT rates, the set $\{k_j\}$, are computed for a hop from NC $i$ to all of its [100] and [111] nearest neighbors, the set $\{j\}$, according to eq. M3. Next, the hopping time to all nearest neighbors, the set $\{t_j\}$, are calculated via inverse transform sampling,

$$t_j = -\frac{\ln[U_j]}{k_j}, \quad (M5)$$

where $U_i$ is a random number pulled (one for each NN $\{j\}$) from a uniform distribution between 0 and 1. The smallest hopping time from the set $\{t_j\}$ is then taken as the hopping time, and the hop occurs to the NC, $j$, which corresponds to this smallest hopping time.

**Superlattice Construction and TOF simulations** Simulations are performed on a BCC superlattice of NCs with unit cell dimensions 32x32x1000. The bandgaps of each NC in the superlattice are pulled from a normal distribution with a mean of 0, and band gap inhomogeneity is characterized by a standard deviation $\sigma_{Eg}$. Deep traps are added to the superlattice structure at random, at a density $\rho_T$, according to the trap depth given by eq. 5 in the main text and eq. M4 in the Supplementary Materials. The levels of each deep traps nearest neighbors are additionally shifted, according to the parameterization presented in Supplementary Materials. We assume that the electrostatic shifts are additive, i.e. two neighboring traps will additionally shift each other's levels further, according to the parameterization in Supplementary Materials.

A single charge carrier is initialized at time $t=0$ on a randomly selected NC at $z=0$. The KMC then proceeds by sequentially hopping the carrier from NC to NC according to the procedure outlined above. The simulation is then terminated once the charge carrier reaches a $z$ value corresponding to the device thickness. The result of the simulation is a set of charge carrier arrival times at a given NC coordinate, $\{t_j, r_j\}$, from which we can compute the TOF measured displacement current,

$$I(t_{j-1} < t \leq t_j) = e \frac{(\vec{r}_j - \vec{r}_{j-1}) \cdot \hat{z}}{(t_j - t_{j-1})d}, \quad (M6)$$

where $d$ is the device thickness. This procedure is carried out stochastically for $10^5$ charge carriers and the results averaged for the overall TOF transient, generating a uniquely disordered NC-superlattice for each charge carrier simulated.

**PbS Synthesis, Device Fabrication and Characterization**

*PbS Synthesis* Colloidal oleic-acid capped PbS NCs are synthesized using the hot injection method. The as synthesized NCs are washed 3 times in mixtures of ethanol and methanol, and finally suspended in hexane at a concentration of 40 mg/mL. We determine the size of the NCs from their absorption spectrum using a well-established parametric model.[4]

*PbS NC-Layer Fabrication* PbS NC-layers were fabricated on substrates described below through sequential dipcoating in (i) PbS NC solution diluted to 5mg/ml in hexane, (ii) crosslinking solution of 6mM ethanedithiol (EDT) in anhydrous acetonitrile, and (iii) rinsing solution of anhydrous acetonitrile. Dip-coating was carried out in air. The thickness of the PbS-NC layers were adjucsted by the number of dip-coating cycles, and thicknesses were measured from SEM cross-sections of the devices after characterization, using ~100 measurements of the thickness over a cross-section spanning the entire device.

***Device Fabrication*** For the standard heterojunction devices, a TiO$_2$ nanoparticle paste (DSL 90-T, Dyesol), diluted to 125mg/mL in acetone was spun on fluorinated tin oxide/glass substrates (Solaronix) at 2500rpm for 60s. Samples were annealed on a hotplate at 500ºC for 60min, then immersed in a 60mM titanium-tetrachloride/deionized water solution at 70 ºC for 30min, and then placed on a hotplate at 500ºC for 60min after thorough rinsing with deionized water. Top contacts of MoOx/Au/Ag (20nm, 100nm, 500nm) electrodes were deposited by thermal evaporation. For the inverted heterojunction devices, NiO was deposited by RF magnetron sputtering using a 99.95% purity NiO target onto indium tin oxide (ITO) glass substrates (Thin Film technologies) using a 10% partial pressure of oxygen. Top contacts of LiF/Al/Ag (10nm, 100nm, 500nm) electrodes were deposited by thermal evaporation.

***Electrical Characterization*** Samples are mounted into a cryostat (Janis ST-500) and remain in vacuum during the measurements. I-V measurements are carried out using a Keithley 2400 source measure unit. A mercury(xenon) DC arc lamp (Newport) and an air mass filter, calibrated using a piezoelectric sensor and an optical power meter (Thor Labs S302C, PM100D) provide AM1.5G illumination for solar cell characterization. For the TOF measurements, the devices are mounted on a Nikon Eclipse Ti-U optical microscope. A 405nm, 100ps excitation pulse is provided by a Hamamatsu picosecond pulsed laser (PLP-10). We note that the excitation energy is larger than the bandgap of the NCs, however, this should not impact the transient dynamics as PbS NCs have hot-carrier cooling times on the order of 100s of fs. Voltage biases were applied using an Agilent 33522A arbitrary waveform generator and the current was measured on a Rohde&Schwarz

RTM1054 oscilloscope through the 50Ω input. Measurements were averaged over 1024 cycles at a frequency of 10kHz.

**Supplementary Information** is linked to the online version of the paper at www.nature.com/nature.

**Acknowledgments:** The authors would like to thank Prof. Jürg Hutter and Dr. Marcella Mauri Iannuzzi for discussions.

**Author contributions:** N.Y. and V.W. devised the work, N.Y. performed all modelling and calculations, with input from S.A., M.H.B.-H., and M.L. for the trap depth calculations, N.Y. fabricated heterojunction devices with materials synthesized by M.Y. and O.Y., using setups maintained by S.V. and W.M.M.L., V.F. and N.Y. fabricated the inverted heterojunction devices, N.Y. performed the TOF experiments and analyzed results, N.Y. and V.W. wrote the manuscript with input from all other authors.

**Funding:** The authors acknowledge an ETH Research Grant (N.Y), the Swiss National Science Foundation Quantum Sciences and Technology NCCR (N.Y.). Computations were supported by grants from the Swiss National Supercomputing Centre (CSCS; project IDs s674, s831, s876).

**Competing Financial interests:** The authors declare that they have no competing interests.

Correspondence and requests for materials should be addressed to vwood@ethz.ch

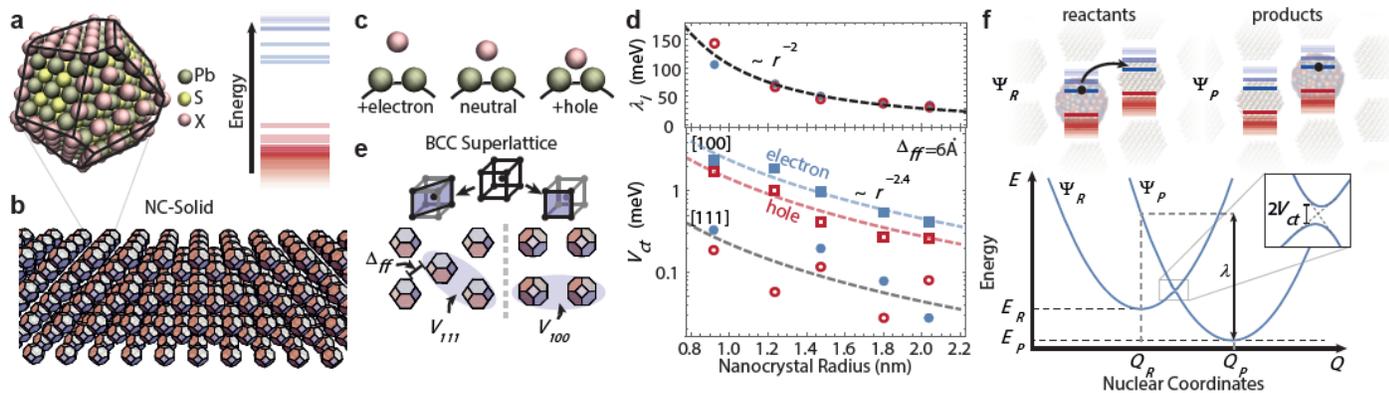

**Fig. 1 Charge Transfer in NC-solids** (**a**) Atomistic model of a PbS nanocrystal (NC) with halide (I) surface passivation and its quantized electronic structure. (**b**) Depiction of a thin film of nanocrystals (i.e, a NC solid). (**c**) Schematic of the nuclear reorganization, where the Pb-X bonds on the surface of the NC expand or shrink in the presence of an electron or hole. (**d**) Calculated reorganization energy (top) and electronic coupling (bottom) for electrons (blue) or holes (red) between [100] (squares) and [111] (circles) nearest neighbors (as depicted in (**e**)), assuming a facet-to-facet distance, $\Delta_{ff}$, of 6 Å in the [111] direction. (**f**) Configurational diagram for charge transfer between two nanocrystals, where an electron (black dot) moves from the nanocrystal on the left (configuration of reactants, $Q_R$, with ground state energy $E_R$) to the nanocrystal on the right ($Q_P$ and energy $E_R$). The reorganization energy ($\lambda$) and electronic coupling ($V_{CT}$) is shown graphically.

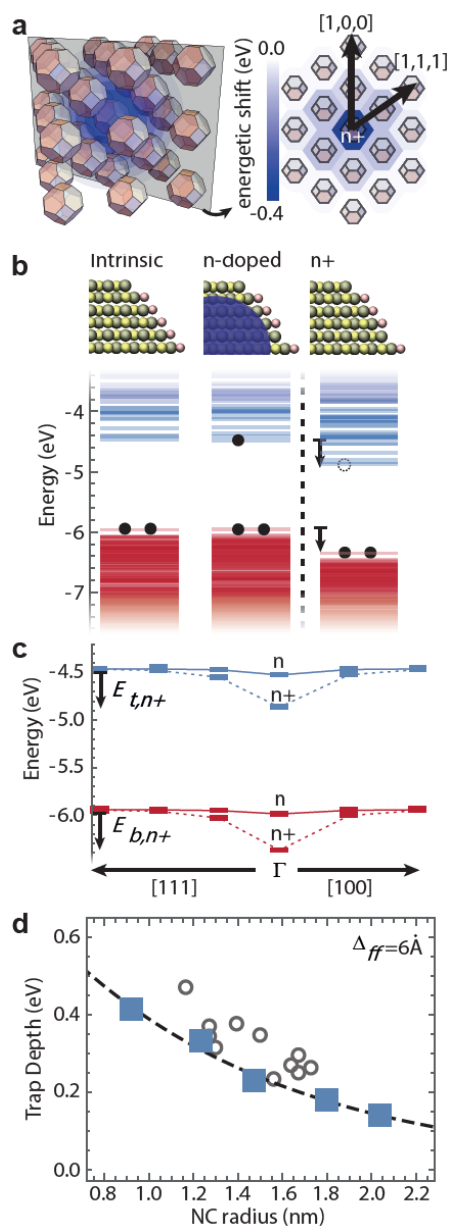

**Fig. 2 Origin of deep traps in NC-semiconductors.** (**a**) A schematic representation of an *n+* or *p-* NC in a solid of intrinsic NCs, where the shift in the energy structure will be screened by neighboring NCs, and a contour plot of the energy shifts of the lowest unoccupied electronic level in NCs ($r = 0.95$ nm with $\Delta_{ff} = 0.6$ Å) for an *n+* NC in an NC solid. (**b**) Electronic structure of intrinsic, n-doped, and oxidized n-doped NC (*n+*). (**c**) Highest occupied electronic levels (red) and lowest unoccupied electronic levels (blue) of an *n*, *n+* NC at the *Γ* point in the BCC lattice and its intrinsic neighbors in the [111] and [100] directions in a NC-solid. (**d**) Trap depth as a function of NC size for NCs in vacuum (circles) and for NC-solids (squares). Experimentally measured trap depths on PbS NC-solids [23,28] (gray circles) and the NC charging energies calculated for a sphere of radius *r* in a PbS NC-solid (dashed gray line) (see **Supplementary Information note 5**).

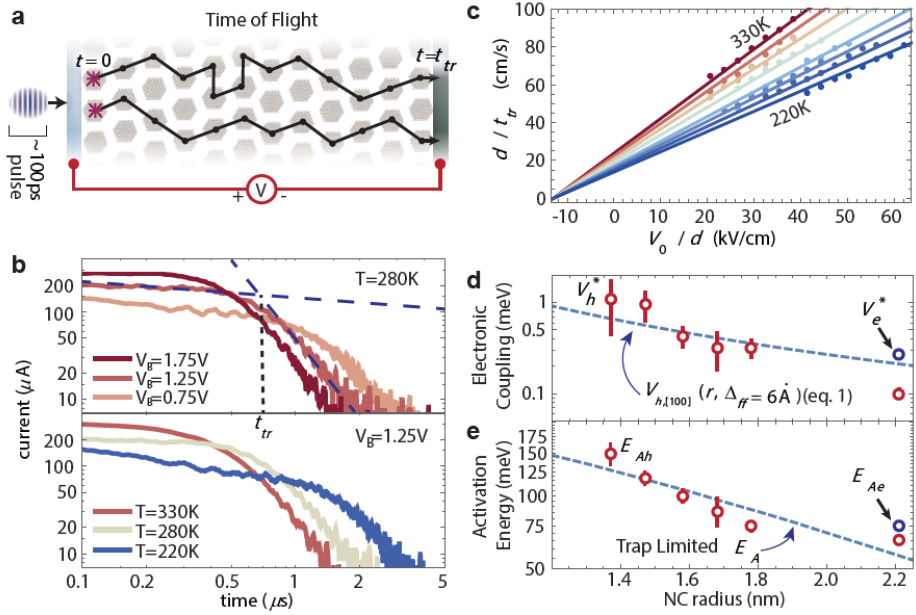

**Fig. 3 Trap-Limited Transport in NC-solid.** (**a**) Schematic of time-of-flight (TOF) photocurrent transient measurements. (**b**) Hole transients measured at various biases and temperatures on a NC-solid ($r_{NC}$ = 2.21 nm). (**c**) Plot of the hole velocity, $d/t_{tr}$, versus applied field, $V_B/d$, at various temperatures. Solid lines indicate the fit to eq. 4. (**d**) Electronic coupling and (**e**) activation energy extracted from TOF measurements as a function of the NC radius $r$ compared to computed $V_{h[100]}$ and $E_A$ assuming trap limited transport (dashed lines).

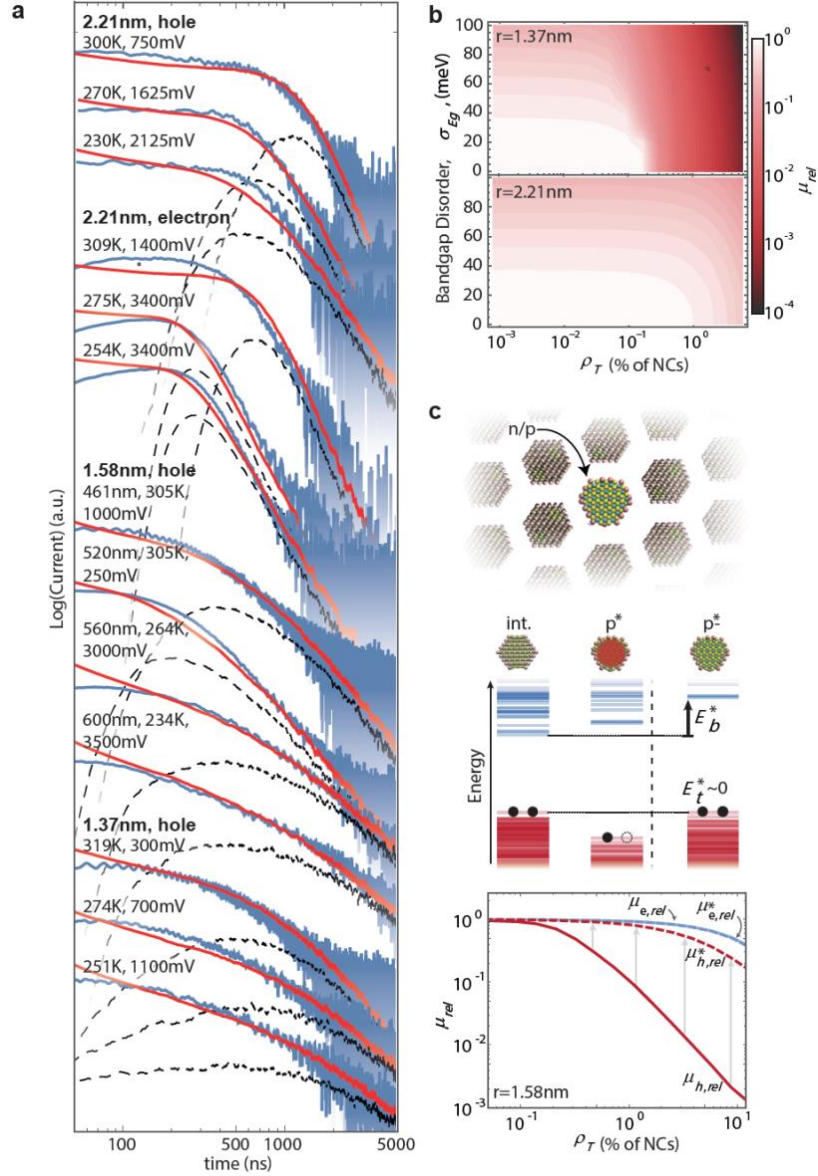

**Fig. 4 Predictive Model for Charge Transport** (**a**) Measured TOF transients (blue) and Kinetic Monte Carlo (KMC) simulated transients (red) are shown for various NC sizes, temperatures, biases, and device thicknesses. Distribution of carrier transit times from the simulated transients (dashed black line). (**b**) Plot of the ratio of the effective mobility as a function of trap density $\rho_T$ and NC bandgap disorder $\sigma_{Eg}$ to the effective mobility of a trap and disorder free NC-solid for smaller (top) and larger (bottom) NC sizes, calculated for a 400nm-thick NC-solid at 300K (**c**) Doping without the formation of trap states can be achieved by introduction of larger-bandgap, *n*- or *p*-doped NCs. Simulations demonstrate that this prevents a decrease in effective mobility at high carrier concentrations.